\documentclass[longauth]{aa}
\usepackage{graphicx}
\usepackage{txfonts}
\usepackage{amsmath}
\usepackage{multirow}
\usepackage{xcolor}
\usepackage[colorlinks=true,linkcolor=purple,citecolor=blue]{hyperref}
\usepackage{url}

\defcitealias{colombo22}{C22}
\defcitealias{Ronchini22}{R22}

\begin{document}

   \title{HERMES: Gamma Ray Burst and Gravitational Wave counterpart hunter}


   \author{G.\ Ghirlanda\fnmsep\thanks{Based on work of the HERMES-Pathfinder collaboration, see list in the Appendix}
          \inst{1,2}
          \and 
          L.~Nava\inst{1}
          \and
          O.~Salafia\inst{1,2}
          \and
          F.~Fiore\inst{3,4}
          \and
          R.~Campana\inst{5,6}
          \and
          R.~Salvaterra\inst{7}
          \and
          A.~Sanna\inst{8}
          \and
          W.~Leone\inst{9}
          \and
          Y.~Evangelista\inst{10,11}
          \and
          G.~Dilillo\inst{10}
          \and
          S.~Puccetti\inst{12}
          \and
          A.~Santangelo\inst{13}
          \and 
          M.~Trenti\inst{14,15}
          \and
          A.~Guzmán\inst{13}
          \and
          P.~Hedderman\inst{13}
          \and
          G.~Amelino-Camelia\inst{16}
          \and
          M.~Barbera\inst{23}
          \and
          G.~Baroni\inst{3}
          \and
          M.~Bechini\inst{17}
          \and
          P.~Bellutti\inst{18}
          \and
          G.~Bertuccio\inst{19}
          \and
          G.~Borghi\inst{18}
          \and
          A.~Brandonisio\inst{17}
          \and
          L.~Burderi\inst{8}
          \and
          C.~Cabras\inst{20}
          \and
          T.~Chen\inst{21}
          \and
          M.~Citossi\inst{3}
          \and
          A.~Colagrossi\inst{17}
          \and
          R.~Crupi\inst{29}
          \and
          F.~De~Cecio\inst{17}
          \and
          I.~Dedolli\inst{19}
          \and
          M.~Del~Santo\inst{22}
          \and
          E.~Demenev\inst{18}
          \and
          T.~Di~Salvo\inst{23}
          \and
          F.~Ficorella\inst{18}
          \and
          D.~Ga{\v c}nik\inst{24}
          \and
          M.~Gandola\inst{19}
          \and
          N.~Gao\inst{21}
          \and
          A.~Gomboc\inst{25}
          \and
          M.~Grassi\inst{26}
          \and
          R.~Iaria\inst{23}
          \and
          G.~La~Rosa\inst{22}
          \and
          U.~Lo Cicero\inst{30}
          \and
          P.~Malcovati\inst{26}
          \and
          A.~Manca\inst{8}
          \and
          E.~J.~Marchesini\inst{5}
          \and
          A.~Maselli\inst{28}
          \and
          F.~Mele\inst{19}
          \and
          P.~Nogara\inst{22}
          \and
          G.~Pepponi\inst{18}
          \and
          M.~Perri\inst{28}
          \and
          A.~Picciotto\inst{18}
          \and
          S.~Pirrotta\inst{12}
          \and
          J.~Prinetto\inst{17}
          \and
          M.~Quirino\inst{17}
          \and
          A.~Riggio\inst{8}
          \and
          J.~\v{R}\'{i}pa\inst{27}
          \and
          F.~Russo\inst{22}
          \and
          D.~Sel{\v c}an\inst{24}
          \and
          S.~Silvestrini\inst{17}
          \and
          G.~Sottile\inst{22}
          \and
          M.~L.~Thomas\inst{14}
          \and
          A.~Tiberia\inst{12}
          \and
          S.~Trevisan\inst{3}
          \and
          I.~Troisi\inst{17}
          \and
          A.~Tsvetkova\inst{8,5}
          \and
          A.~Vacchi\inst{29}
          \and 
          N.~Werner\inst{27}
          \and
          G.~Zanotti\inst{17}
          \and
          N.~Zorzi\inst{18}
}

   \institute{INAF -- Osservatorio Astronomico di Brera,
              via E. Bianchi 46, I-23807 Merate (LC), Italy
          \and
              INFN -- Sezione di Milano-Bicocca, piazza della Scienza 3, I-20126 Milano, Italy
          \and
         INAF -- Osservatorio Astronomico di Trieste, 
         Via G.B. Tiepolo 11, I–34143 Trieste, Italy.
         \and
         IFPU -- Institut for Fundamental Physics of the Universe, Via Beirut 2, I-34014 Trieste, Italy.
         \and
         INAF -- OAS, Via Piero Gobetti 101, Bologna, I-40129, Italy.
         \and
          INFN --- Sezione di Bologna, Viale Berti Pichat 6/2, Bologna, I-40127, Italy
         \and
         INAF -- Istituto di Astrofisica Spaziale e Fisica Cosmica di Milano, Via A. Corti 12, 20133 Milano, Italy
         \and 
         Dipartimento di Fisica, Università degli Studi di Cagliari, SP Monserrato--Sestu km 0.7, 09042 Monserrato, Italy.
         \and
         Dipartimento di Fisica, Università di Trento, Sommarive 14, 38122 Povo (TN), Italy
         \and
         INAF -- Istituto di Astrofisica e Planetologia Spaziali, Via del Fosso del Cavaliere 100, I-00133 Roma (RM), Italy.
         \and
         INFN -- Sezione di Roma Tor Vergata, Via della Ricerca Scientifica 1, I-00133 Roma (RM), Italy
         \and
         Agenzia Spaziale Italiana, Via del Politecnico snc, 00133 Roma, Italy
         \and
         Institut für Astronomie und Astrophysik, Universität Tübingen,
        Sand 1, D-72076 Tübingen, Germany
        \and
        School of Physics, University of Melbourne, Parkville, Vic 3010, Australia
        \and
        Australian Research Council Centre of Excellence for All-Sky Astrophysics in 3-Dimensions, Australia
        \and
        Physics Department, Federico II University, via Cintia 21, I-80126, Napoli, Italy
        \and
        Politecnico di Milano - Department of Aerospace Science and Technology, Milan, Italy
        \and
        Fondazione Bruno Kessler, Center for Sensors and Devices, Trento, Italy
        \and
        Politecnico di Milano - Department of Electronics, Information and Bioengineering, Como, Italy
        \and
        Dipartimento di Matematica e Informatica, Università degli Studi di Cagliari, SP Monserrato--Sestu km 0.7, 09042 Monserrato, Italy.
        \and
        Key Laboratory of Particle Astrophysics, Institute of High Energy Physics, Chinese Academy of Sciences, Beijing, 100049, China
        \and
        INAF - IASF Palermo, Via Ugo La Malfa 153, 90146 Palermo, Italy
        \and
        Dipartimento di Fisica e Chimica, Universita` degli Studi di Palermo, via Archirafi 36, I-90123 Palermo, Italy
        \and
        SkyLabs d.o.o., Zagreb{\v s}ka c. 104, 2000 Maribor, Slovenia
        \and
        Center for Astrophysics and Cosmology, University of Nova Gorica, Vipavska 13, 5000 Nova Gorica, Slovenia
        \and
        Department of Electrical, Computer and Biomedical Engineering, University of Pavia, Italy
        \and
        Department of Theoretical Physics and Astrophysics, Faculty of Science, Masaryk University, Brno, Czech Republic
        \and
        INAF – Osservatorio Astronomico di Roma, Via Frascati 33, I–00078 Monte Porzio Catone (RM), Italy
        \and
        Department of Mathematical, Informatics, and Physics University of Udine
        \and
        INAF - OAPA, piazza del Parlamento 1, 90134 Palermo, Italy\\
\email{giancarlo.ghirlanda@inaf.it}
             }

   \date{Received ...; accepted ...}

 
  \abstract
  {Gamma Ray Bursts (GRBs) bridge relativistic astrophysics and multi--messenger astronomy. Space--based $\gamma$/X--ray wide field detectors have proven essential to detect and localize the highly variable GRB prompt emission, which is also a counterpart of gravitational wave events. We study the capabilities to detect long and short GRBs by the \emph{High Energy Rapid Modular Ensemble of Satellites} (HERMES) Pathfinder (HP) and SpIRIT, namely a swarm of six 3U CubeSats to be launched in early 2025, and a 6U CubeSat launched on December 1st 2023. We also study the capabilities of two advanced  configurations of swarms of $>$8 satellites  with improved detector performances (HERMES  Constellations). The HERMES detectors, sensitive down to $\sim$\,2-3\,keV, will be able to detect faint/soft GRBs which comprise X-ray flashes and high redshift bursts. By combining state-of-the-art long and short GRB population models with a description of the single module performance, we estimate that HP will detect $\sim$\,$195^{+22}_{-21}$ long GRBs ($3.4^{+0.3}_{-0.8}$ at redshift $z>6$) and $\sim$\,$19^{+5}_{-3}$ short GRBs per year. The larger HERMES Constellations under study can detect between $\sim$\,1300 and $\sim$\,3000 long GRBs per year and between $\sim$\,160 and $\sim$\,400 short GRBs per year, depending on the chosen configuration, with a rate of long GRBs above $z>6$ between 30 and 75 per year. Finally, we explore the capabilities of HERMES to detect short GRBs as electromagnetic counterparts of binary neutron star (BNS) mergers detected as gravitational signals by current and future ground--based interferometers. Under the assumption that the GRB jets are structured, we estimate that HP can provide up to $\sim$1  (14) yr$^{-1}$ joint detections during the fifth LIGO-Virgo-KAGRA observing run (Einstein Telescope single triangle 10--km--arm configuration). These numbers become $\sim$\,4 (100) yr$^{-1}$, respectively, for the HERMES Constellation configuration.  }

   \keywords{gamma-ray bursts --
                X-rays --
                gravitational waves --
                gamma-ray detectors
               }

   \maketitle
%

\section{Introduction}
Gamma-ray bursts (GRBs) release a large amount of electromagnetic energy ($E_\gamma$\,$\sim$\,$10^{51}$\,erg) over a short timescale ($\sim$\,$10^{-2}-10^3$\,seconds) as a result of energy dissipation in a strongly collimated (a few degrees wide) jet, in which particles move at ultra-relativistic velocities ($\Gamma\gtrsim100$).
The initial radiative phase (i. e., the `prompt emission') is detected in the keV-MeV range. The flux variability time scale during such phase can be as short as 1--10\,ms \citep{maclachlan13,Golkhou2015}. The origin of this fast variability, its connection to the central engine, to the properties of the region where the radiation is produced, and the origin of the radiation itself are still largely unknown.

Traditionally, GRBs have been categorized into `long' and `short' bursts based on their observed prompt emission duration, with a separation\footnote{Most recent results have shown that the separation in observer frame duration does not map exactly the two progenitor channels, see e.g. GRB~211211A \citep{Rastinejad2022} and GRB~200826A \citep{Bromberg2013,Rossi2022}.} at 2 seconds \citep{Kouveliotou1993}. Direct (imaging) and indirect (light curve photometry and spectroscopy) long GRB-supernova associations have now established that many long-duration GRBs originate from the core-collapse of massive stars \citep{Leva2016}, as originally hypothesized by \citet{Woosley1993}.  The first conclusive evidence linking short-duration GRBs to the merger of compact object binaries including at least one neutron star, as initially proposed by \citet{Eichler1989} and \citet{Narayan1992}, came from the association of GRB170817A \citep{Abbott2017} and the kilonova AT2017gfo \citep{coulter17,pian17} with the first gravitational wave (GW) signal from the merger of two neutron stars \citep{Abbott2017ApJ} detected by the Advanced Laser Interferometer Gravitational wave Observatory (aLIGO, \citealt{Aasi2015}) and Advanced Virgo \citep{Acernese2015}.

The detection rates of compact binary mergers will increase in the next decade, thanks to joint observation runs of the Advanced Virgo, aLIGO and KAGRA \citep{Akutsu2020}. Predictions of the joint electromagnetic (EM) and GW detections are currently under development (see e.g. \citealt{colombo22,Colombo2023}). A major step forward in the detection capabilities of GWs will be achieved with third generation interferometers, such as the Einstein Telescope (ET --  \citealt{Punturo2010,maggiore20,Branchesi2023}) and the Cosmic Explorer (CE -- \citealt{Abbott2017,Reitze2019}) which will enable the discovery and follow-up of hundreds of EM counterparts (see e.g.\ \citealt{Ronchini22}).

The GRB prompt emission is followed by a longer-lasting afterglow emission, produced in the interaction of the jet with the external medium. This component is typically detected in the X-rays and optical bands and, less often, in the radio and in the GeV gamma-rays. Starting from 2018, five GRB afterglows have been firmly detected at very-high energies (VHE, $>$100\,GeV) as well (for a recent review, see \citealt{miceli22}). Photons with energies up to $\sim$\,10\,TeV have been detected by LHAASO from GRB\,221009A \citep{221009a_lhaaso}, the brightest GRB ever detected. There is a general consensus in identifying the origin of this emission as synchrotron self Compton component \citep{190114c_interp,salafia22b}, although other possibilities are being investigated \citep{190829a}. The detections at VHE are still very limited and sparse, and a full exploitation of VHE observations will be achieved \citep{Bernardini2019} only with the advent of the Cherenkov Telescope Array (CTA), the next generation of imaging atmospheric Cherenkov telescopes (IACT). The full array is expected to be operational from 2025. Similarly, at the other extreme, in the radio band, the follow up and imaging of GRB jets have demonstrated to be a unique opportunity to dig into the jet (e.g. \citealt{Mooley2018,Ghirlanda19}) and ambient medium properties (e.g. \citealt{Salafia2022,Giarratana2022}). The current limited number of radio detected GRBs \citep{Chandra2012} should increase with the advent of the Square Kilometer Array (SKA) and its pathfinders \citep{Ghirlanda2013,Ghirlanda2014}.

The lesson learned over the past 50 years, since the discovery of GRBs, underscores that almost all the GRB-related science, including the recent multi-messenger (MM) advancements, is enabled by the detection and localisation of the GRB prompt event. 
Currently, the most active space telescopes for the search and study of GRBs are the Neil Gehrels {\it Swift} Observatory \citep{Gehrels2004} and the {\it Fermi} satellite \citep{Meegan2009}, operational since 19 and 15 years, respectively. Together they detect about 300\,GRBs per year. The continuation of their activity after 2024 is not guaranteed. 
{\it Swift}--BAT (15--350 keV) provides good GRB localisations (several arcminutes) enabling rapid follow up in other bands, eventually leading (in about 30\% of events) to the determination of the redshift of the GRB. On the other hand, {\it Fermi} (8 keV -- 100 GeV), while providing less precise localisations (to within several degrees), allows the study of the temporal and spectral variability of prompt emission across nearly seven orders of magnitude in energy.

In order to maximise the impact that major future facilities, such as CTA, SKA  and next generation gravitational wave interferometers, may have in the GRB and multi-messenger astronomy related fields, a facility hunting GRBs and providing accurate localisations and temporal/spectral information is needed.

At the present stage, the Space Variable Objects Monitor (SVOM, \citealt{Atteia2022}) is planned to become operative by the end of 2024 and the Einstein Probe satellite (EP, \citealt{Yuan2022}) has been successfully launched on January 2024. The SVOM estimated rate of alerts with localisation capabilities similar to BAT is 50--60 GRB\,yr$^{-1}$, while for additional 90\,GRBs the localisation is expected to be $\lesssim5-10$\,degrees. EP can provide several GRB detections with accurate localisations for multi-band follow up of their afterglow emission. However, its limited energy range (0.3--5 keV for the Wide X-ray Telescope extended up to 10 keV by the narrow field Follow up X--ray Telescope) prevents a detailed spectral characterisation of the prompt GRB emission as currently possible with e.g. {\it Fermi}. 

Over the next couple of decades, mission concepts such as the Transient High-Energy Sky and Early Universe Surveyor (THESEUS, \citealt{Amati2021}) or the Gamow Explorer \citep{White2021} are under study. THESEUS, comprising a suite of instruments sensitive in the keV--MeV energy range with imaging capabilities at the lower end of this range, will detect hundreds of GRBs per year, with a considerable fraction detected at high redshifts. The fast repointing capabilities and the presence of an infrared telescope onboard will secure an increase by a factor of 10 of the current eight GRBs with measured distance at $z\ge6$ \citep{Amati2021}.  

Given this status of the field, it would be highly beneficial to complement existing and planned observatories through a technology which can be built on short time scales (a few years) and at an affordable cost \citep{Fiore2021}. The High Energy Rapid Modular Ensemble of Satellites (HERMES) discussed in this paper is conceived to serve this purpose. GRB monitoring by CubeSats has been successfully demonstrated by e.g. GRBalpha \citep{Pal2020,Pal2023,Ripa2022} and VZLUSAT-2 \citep{Granja2022} which have detected $\sim$60 and $\sim$30 GRBs up to February 2024. GRBAlpha is a 1U CubeSat, launched on March 22nd 2021 on a Low Earth Sun Synchronous orbit (SSO). It hosts a CsI(Tl) scintillator crystal read out by an array of Silicon Photo Multipliers providing a peak collecting area of $\sim50~$cm$^2$ in the 70--800 keV energy band. VZLUSAT-2 was launched in an SSO on January 13th, 2022 and hosts two gamma-ray detectors similar to that on board GRBAlpha. GRBAlpha and VZLUSAT-2 are finishing or surpassing their third and second year in orbit, showing the sustained-in orbit reliability achieved by CubeSats and instruments for high-energy astrophysics. 

In this study, we calculate the detection rates of short and long GRBs by HERMES. We derive the characteristics of GRBs that can be detected by HERMES and compare them with existing datasets of GRBs detected by {\it Swift}/BAT and {\it Fermi}/GBM. Additionally, we assess the rate of joint detection of short GRBs by HERMES and GWs from  binary neutron star mergers by aLIGO and Advanced Virgo in O5, and by ET at design sensitivity.

The paper is organised as follows. Section 2 introduces the HERMES mission. In Section 3, we evaluate the capabilities of HERMES in detecting both short and long GRBs. Section 4 focuses on the joint EM-GW detections with ET. We provide a discussion and summary of the findings in Section 5. Throughout the paper we adopt a standard flat cosmology $\Omega_M=0.3, h=0.7$.

\section{The HERMES mission}

HERMES\footnote{\url{https://www.hermes-sp.eu}} is a modular X-ray space mission specifically dedicated to the study of GRBs and fast X-ray transients. To demonstrate its capabilities, three technological (HERMES-TP, funded by the Italian Space Agency, ASI) and three scientific (HERMES-SP, funded by the European Commission Horizon 2020 Research and Innovation Program) pathfinder units are in preparation. Together, all these constitute the HERMES Pathfinder (HP hereafter). The primary goals of the HP are the validation of the capability to detect GRBs and other fast transients with miniaturized instrumentation hosted by CubeSats and to obtain their localisation using the triangulation technique \citep{Sanna2023}, which measures the arrival time delay of the signal across different detectors located thousands of kilometres apart \citep{Fiore2020,Fiore2022}. 

The HERMES Pathfinder is composed by six units, each measuring 10$\times$10$\times$30 cm for a total weight of 6 kg, classified as 3U CubeSats. These units are scheduled for launch in early 2025. Additionally, a seventh instrument payload, identical to those used in HP, is hosted by the SpIRIT (\emph{Space Industry Responsive Intelligent Thermal nanosatellite}\footnote{\url{https://spirit.research.unimelb.edu.au}}) CubeSat, led by the University of Melbourne \citep{Trenti2021}. SpIRIT, a 6U CubeSat, has been launched in SSO on December 1st, 2023. It is currently in a commissioning phase and scientific observations are expected to start in the second quarter of 2024.
So far its performance appears nominal. This crucial experiment is validating the instrument design, completely realised by INAF and its partner Politecnico di Milano. The approach chosen for the development of the instrument is based on the design and realisation of critical sub-systems such as for example the silicon drift detectors (SDD) and the on board software, on the basis of scientific agreements with the providers Fondazione Bruno Kessler and University of Tubingen, and the use of COTS components. SpiRIT will also be crucial to test the HP payload in the harsh space environment and in different thermal conditions. The SDDs employed by the HP payload suffer damage due to energetic particles such as electrons and protons. SpIRIT will allow us to quantify for the first time the amount of the degradation of SDD performances in SSO. The SDD noise properties depends also on the operating temperature, improving toward lower temperatures. Among the SpIRIT payloads there is also an active temperature control system, based on a Stirling cycle cryocooler and deployable thermal radiators (the TheMIS payload). We will therefore test the performances of the HP payload in a wide temperature range, from $-$30~$^\circ$C (the operational limit of the HERMES payload) to +20~$^\circ$C. This will allow us to identify the optimal temperature range for using the instrument, and therefore the requirements for the thermal control system in view of the application on the HERMES Constellation.

Therefore, the HERMES Pathfinder in conjunction with SpIRIT will consists of seven satellites on two different orbital planes. SpIRIT and HERMES Pathfinder serve as a test for the project of the a larger constellation comprising tens of CubeSat modules (HERMES Constellation, hereafter).

HERMES Pathfinder and SpIRIT host simple but innovative X-ray detectors, characterised by a particularly wide energy band, from $2-3$\,keV to $\sim2$\,MeV, and an excellent temporal resolution of about 300\,ns, three to seven times better that the best resolution achieved so far for GRB studies.
Each CubeSat hosts a detector (see e.g.\, \citealt{Evangelista2020,Evangelista2022}) sensitive to both X-rays (X-mode or simply X, hereafter) and soft $\gamma$-rays (S-mode or S, hereafter). The XGIS instrument on board THESEUS \citep{Amati2021} is based on the same ``siswich'' architecture. The characteristics of each HP unit are listed in the first row of Table~\ref{tab:instrument_parameters}. The effective area as a function of photon energy is shown in Fig.~\ref{fig:eff_areas} for different values of the boresight angle $\theta$ between the source direction and the detector normal (see \citealt{campana20,campana2022} for more details).  The shaded regions in Fig.\ref{fig:eff_areas} show the energy ranges (also reported in Table~\ref{tab:instrument_parameters}) used for the computation of the detection rate (\S2). The effective area of the detector decreases with increasing boresight angle, especially for the lower photon energies for which we plot only the effective area up to 60 degrees.
The effective area shown in  Fig.\ref{fig:eff_areas} were computed by means of Geant4-based Monte Carlo simulations \citep{agostinelli03} using an updated satellite mass model (much more accurate with respect to the simplified version discussed in \citealt{campana20}). The mass model includes a detailed geometrical and physical description of the detector and of the main components of the spacecraft avionics.
For simplicity, the small dependence over the azimuth angle for fixed $\theta$ values has been here averaged out.
Moreover, the background count rates reported in Table~\ref{tab:instrument_parameters} have been estimated by simulating all the various expected contributions in LEO (e.g., CXB, cosmic rays, Earth albedo, etc. See \citealt{campana13} for more details on the assumed models).
HERMES orbital configuration foresees the satellites to be released in very close orbits that, however, guarantees a natural drift along the velocity-bar direction to achieve the baseline distance between the elements of the constellation \citep{Colagrossi2020}. The HERMES  natural dynamics evolves in a periodic motion of about 80 to 100 d, and the pointing direction of each spacecraft shall be controlled and optimized to cope with this dynamical evolution [2]. Then, to maintain at least three spacecraft's FOV overlapped, and to maximize the sky coverage, an optimized pointing strategy will be commanded to the spacecraft according to a dedicated pointing optimization software \citep{Colagrossi2020,colagrossi2019semi}.

The full HERMES Constellation is based on the HP concept, but it is planned to carry more sensitive detectors onboard. We carry out here a preparatory study by investigating two different configurations for the detectors onboard the full constellation. In particular, we consider detectors with a larger (by a factor of 4) effective area than HP and a field of view similar to that of the HP (Constellation A) or smaller by a factor 2 (Constellation B). The latter assumption of a smaller FoV ensures that Constellation B has a lower X-ray background rate than Constellation A. The properties of a single unit for the Constellation A and B are listed in Table~\ref{tab:instrument_parameters}. 
For the configuration called Constellation A we assume 9 CubeSats simultaneously in orbit, while for Constellation B we assume 13 CubeSats to compensate for their smaller field of view and assure a full sky coverage. The sky coverage maps (Fig.~\ref{fig:mappe}) show for each location in the sky the minimum boresight angle among all the detectors observing that location.

\begin{table*}
    \centering
    \begin{tabular}{|c|c|c|c|c|c|}
    \hline
 &   Detection & FoV  & Energy range  & Back. rate & Effective  \\
 &   mode      & sr   &    keV        &  counts/s  &  area         \\
    \hline
\multirow{2}{*}{HP}  & X    & 3.14 & 3--20  & 352  &  nominal (Fig.~\ref{fig:eff_areas})   \\
    & S    & 5.20 & 50--500        & 80   &  nominal (Fig.~\ref{fig:eff_areas})   \\
    \hline
\multirow{2}{*}{Const. A} & X        & 3.14 & 3--20          & 1408   &   nominal $\times$\,4   \\
      & S        & 5.20 & 50--500        & 320    &   nominal $\times$\,4   \\
    \hline
\multirow{2}{*}{Const. B} & X        & 1.57 & 3--20          & 704  &   nominal $\times$\,4    \\
       & S        & 2.60 & 50--500        & 160   &   nominal $\times$\,4    \\
    \hline
    \end{tabular}
    \vskip 0.2cm
    \caption{Characteristics describing one single HERMES module. For the HP module and the two different configurations of the Constellation (A and B) and for each detection mode (X and S), the table reports the field-of-view (FoV) of a single unit, defined as the field corresponding to 20\% of the on-axis collecting area at 100 keV, the energy range considered for the computation of GRB detection rates, the expected background count rate in the same energy range, and the effective area. For the expected background and the effective areas, see also \cite{campana20,campana2022}.
    }
    \label{tab:instrument_parameters}
\end{table*}

\section{GRB detection rates}
\begin{figure*}
\centering
{\includegraphics[width=9cm]{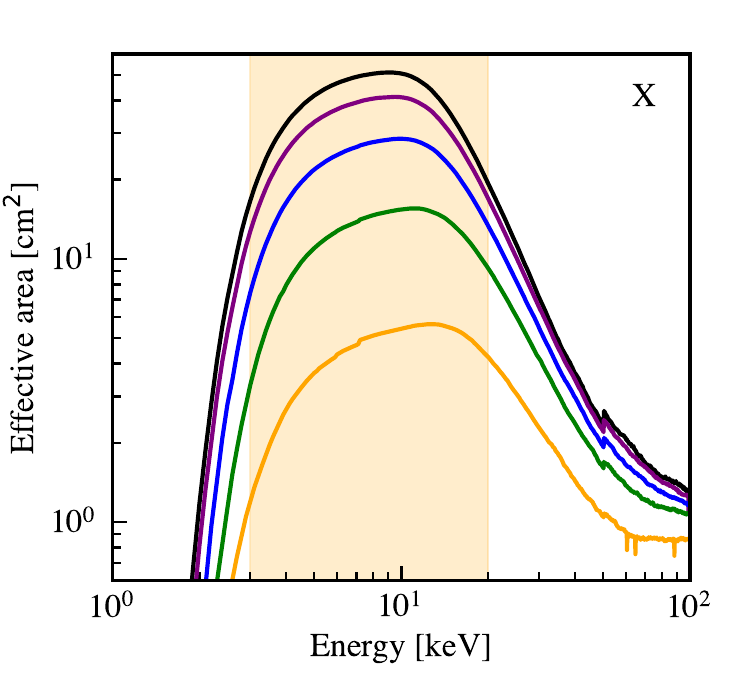}
\includegraphics[width=9cm]{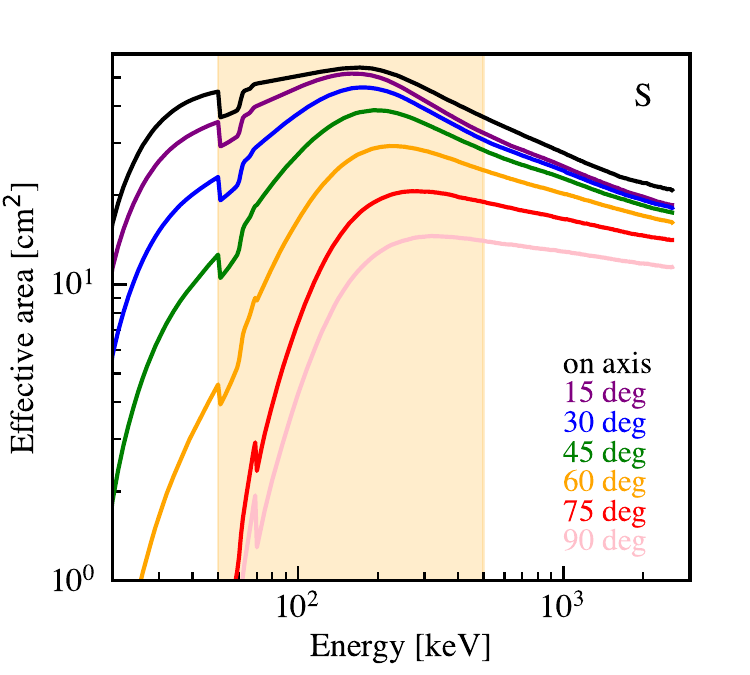}}
\caption{Effective areas of the X-mode (left-hand panel) and the S-mode (right-hand panel) for the HERMES Pathfinder.Different color lines show the effective areas as a function of the boresight angle (as reported in the legend). For the X-mode at angles larger than 60 degrees the effective area degradetes substantially and is not shown for clarity. Shaded vertical stripes show the energy ranges considered in this study to estimate the detection rates. The full HERMES Constellation is assumed to host detectors with a four times larger effective area. }
\label{fig:eff_areas}
\end{figure*}

We employ a population-synthesis approach to assess the GRB detection capabilities of HERMES. This involves creating a realistic mock population of GRBs through Monte Carlo simulations. For each simulated GRB, we calculate its likelihood of being detected by HERMES, taking into account the instrumental characteristics outlined in Table~\ref{tab:instrument_parameters} and illustrated in Fig.~\ref{fig:eff_areas}. This allows for a more detailed analysis of the instrumental performance, such as the boresight-angle dependence of the effective area (Fig.~\ref{fig:eff_areas}) or the sky coverage of a swarm of satellites (Fig.~\ref{fig:mappe}), besides simple scaling of factors like field of view or duty cycle. 

We adopt state-of-the-art population models to describe short and long GRBs. The reliability of the population model is ensured by its calibration with actual GRB datasets (e.g. the GRBs detected by {\it Fermi} and {\it Swift}). For long GRBs, our model is based on the work of \citet{ghirlanda15}. For short GRBs, we employ the most recent model developed by \citet{Salafia2023}. This model provides a comprehensive description of the short GRB population within the framework of a quasi-universal structured jet scenario, as motivated by the expectations that jet should feature an angular structure, and corroborated by the modelling of the multi-wavelength afterglow observations of GRB~170817A (see e.g.\ \citealt{Salafia2022} and references therein). 

\subsection{Detection probability estimate}

For HP and SpIRIT, we first consider one single module and assign to each mock GRB an angle $\theta$ with respect to the detector normal by distributing the GRBs isotropically in the sky. For the X detector, the effective area becomes negligible for boresight angles $\theta\gtrsim 60^{\circ}$ so that any mock GRB located at large angles is considered undetected by the X--mode.  Despite the effective area of the S--mode is not so small at 90 degrees, we consider that possible effects such as the absorption of X--ray photons by the Earth atmosphere may limit the FoV of the S--mode to $\theta<80^{\circ}$. 
For Constellation A and B, we use the orbital sky coverage maps (Fig.~\ref{fig:mappe}) to identify for each mock GRB the minimum boresight angle among the detectors.

We estimate the source photon counts in the detector over a given energy range $E_1$--$E_2$ by integrating the source spectrum $N(E)=dN/dA\, dE\,$ multiplied by the effective area $A_{\rm eff}(E,\theta)$ estimated at the angle $\theta$:
\begin{equation}
    S(\theta)=\int_{E_1}^{E_2} N(E) A(E,\theta) dE.
    \label{Eq:source}
\end{equation}
Here $N(E)$ is the observer-frame photon spectrum integrated over a given time bin $\Delta t$. 
The integral in Eq.~\ref{Eq:source} is performed over the 3--20\,keV energy range for the X-mode (shaded vertical stripe in Fig.~\ref{fig:eff_areas}, left panel) and 50--500\,keV for the S-mode (shaded vertical stripe in Fig.~\ref{fig:eff_areas}, right panel). The expected average background count rates (counts/s) over the same energy ranges are given in Table~\ref{tab:instrument_parameters} \citep{campana20}. The total background counts $\mathcal{B}$ are then obtained by multiplying the background count rate by the same duration $\Delta t$ over which the source spectrum has been accumulated.

The detection probability is determined by the comparison between the number of expected background counts $\mathcal{B}$ and the measured counts $\mathcal{C}$ (which we set, for simplicity and given the small Poisson scatter, to the floor of the sum of the expected source counts $\mathcal{S}$ and background counts $\mathcal{B}$). We consider a GRB as detected if the probability associated with measuring at least $\mathcal{C}$ counts when the expected counts are $\mathcal{B}$ corresponds to detection at more than $5\,\sigma$, that is
\begin{equation}
 P_{\rm det}=\sum_{\mathcal{N}=\mathcal{C}}^{\infty}p(\mathcal{N}|\mathcal{B})=1-\sum_{\mathcal{N}=0}^{\mathcal{C}-1}p(\mathcal{N}|\mathcal{B})\le 2.9\times10^{-7},
\end{equation} 
where
\begin{equation}
    p(\mathcal{N}|\mathcal{B})=\frac{\mathcal{B}^\mathcal{N}\exp{(-\mathcal{B})}}{\mathcal{N!}}.
    \label{eq:prob}
\end{equation}

While the HERMES trigger algorithms under study will benefit from new approaches (see \citealt{Dilillo2023}, \citealt{ward2023poisson} and \citealt{Crupi2023}), a typical generic trigger algorithm is based on the continuous search for a significant increase of the count rate with respect to the average background counts. The current HERMES algorithm is similar to this latter case since it uses a Bayesian approach that looks at the combined likelihood of signals in the detector to alert of potential transients or GRBs \citep{Guzman2020, Belanger2013}. Without entering into the details, which also depend on the light curve shape (e.g. \citealt{Lien14}), we consider two extreme trigger criteria, one based on the peak flux and the other one based on the fluence.

For the trigger criterium based on the fluence, the source counts $\mathcal{S}$ are estimated by considering the total emission over the burst duration $\Delta t=T_{\rm dur}$. The latter is computed for long GRBs as $\Delta t\sim (1+z) E_{\rm iso}/L_{\rm iso}$ while for short GRBs, we assign the durations randomly from a log-normal distribution centered at 0.3 seconds with 0.2 dex dispersion.  The background counts are estimated by multiplying the average background count rate by the same duration $T_{\rm dur}$. If the burst duration is longer than 128\,s, the fluence for the trigger condition is accumulated only up to 128\,s to mimic the anticipated trigger algorithms that could be implemented for HERMES. 

For the trigger criterium based on the peak flux, the source counts are obtained by multiplying the GRB peak flux by the duration over which the peak flux has been estimated. For long GRBs, the duration is $\Delta t=$ 1\,s, that is, we assume that the peak flux is representative of the average flux in the brightest 1\,s time bin. For short GRBs, peak fluxes are estimated on a time bin of 64\,ms. In both cases, if the time bin is longer than the burst duration, only the trigger criterium based on the fluence is considered. The background counts are estimated by multiplying the background count rate (in counts\,s$^{-1}$, see Table~\ref{tab:instrument_parameters}) by the same duration interval $\Delta t$.

\subsection{Results}
We apply the above methods to the HP and to Constellations A and B. As summarised in Table~\ref{tab:instrument_parameters}, the Constellation detectors have a 4-times larger effective area than HP. The difference between configuration A and B of the Constellation is the single detector FoV, that is, equal to that of the HP single module for Constellation A and a factor of 2 smaller for Constellation B (Table~\ref{tab:instrument_parameters}). The reason for considering a smaller FoV is to limit the X-ray background, which is dominated by the cosmic X-ray background and is approximately proportional to the instrument FoV. 

For the HP, we first estimate the detection rate of one single satellite, assuming a duty cycle of 50\%. This is a conservative assumption, and a larger duty cycle is expected in the case of an equatorial orbit. Then, we multiply the detection rate by a factor of 2, assuming that the six satellites are composed of two triplets observing different regions of the sky. In other words, each group of three satellites covers a field of view of $\sim$5.20 sr (with the S-mode instrument) and $\sim$3.14 sr (with the X-mode instrument).

\begin{figure*}
\centering
{\includegraphics[width=8.8cm]{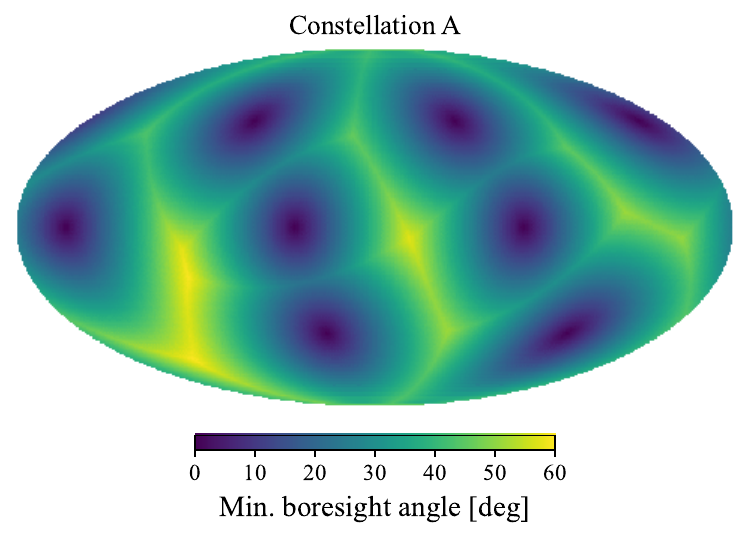}
\includegraphics[width=8.8cm]{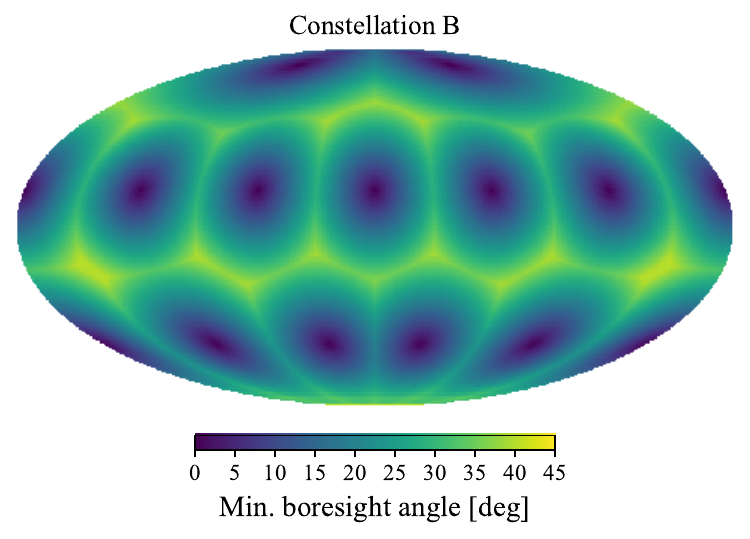}}
\caption{The maps show the minimum angle between the detector normal and the GRB location for a constellation of 9 cubesats (Constellation A - left) and 13 cubesats (Constellation B - right).}
\label{fig:mappe}
\end{figure*}

For Constellations A and B, we provide the detection rate assuming the simultaneous use of a sufficient number of satellites pointing in different directions and covering the full sky. The larger the number of satellites in the constellation, the larger will be the overlap of their fields of view (each satellite of the constellation has a field of view as reported in Table~\ref{tab:instrument_parameters}). We assume a total of 9 units for Constellation A and 13 for Constellation B. The coverage maps for the two configurations are shown in Fig.~\ref{fig:mappe}. The all-sky coverage contributes, together with the increased sensitivity of the single module (owing to the larger effective area), to the increase of the detection rates by the constellation configurations. 

We consider a GRB detected by HERMES if observed in at least one of the two detection modes (X or S) and by at least one of the two trigger algorithms (peak flux or fluence). 
In order to derive the detection rates and their associated uncertainties, we simulate 500 long and 500 short GRB populations by sampling the posterior distributions of the parameters defining the populations. We estimate the detection rates for every single realization of the population and derive, as final detection rates, the 50th percentile of the 500 rate values distribution. 
Uncertainties at the 68\% credible interval are estimated taking the 16th and the 84th percentiles.
The inferred rates and their uncertainties\footnote{Note that as these rates are estimated independently as the 50th percentile of the 500 rate values distribution, the total values reported in Table~\ref{tab:detection_rates} are not strictly the sum of columns 3 and 4.} are reported in Table~\ref{tab:detection_rates}. The table shows, both for short and long GRBs, the total rate (number of detected GRBs per year) and also partial rates: all the GRBs detected by S, by X, and by X-only.  

\begin{figure*}
{
\hspace{0.4 truecm}
\includegraphics[scale=0.54]{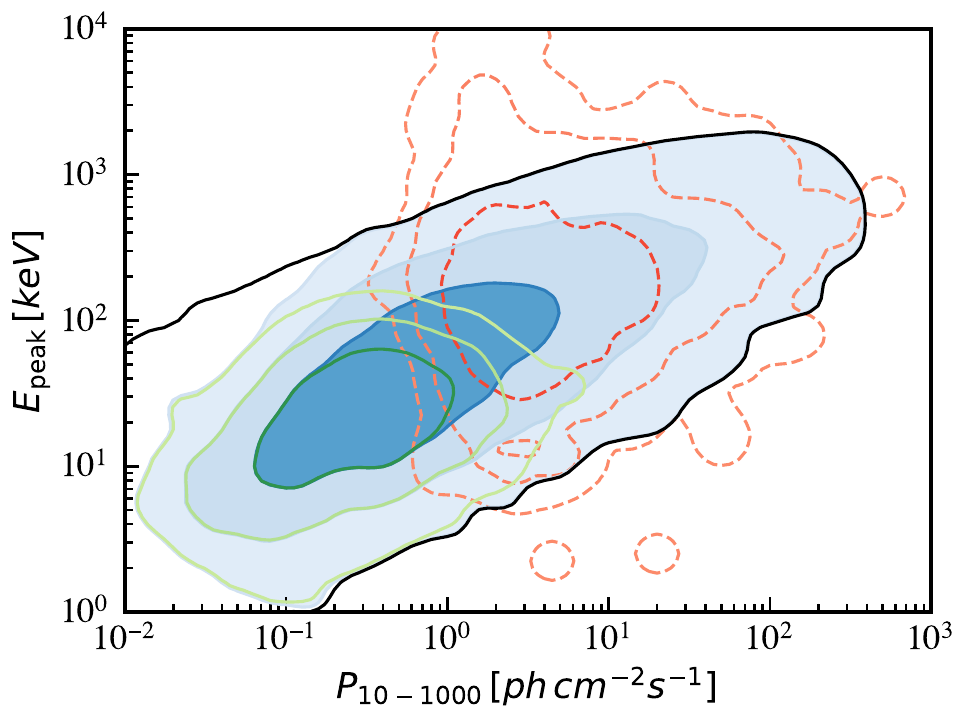}
\includegraphics[scale=0.54]{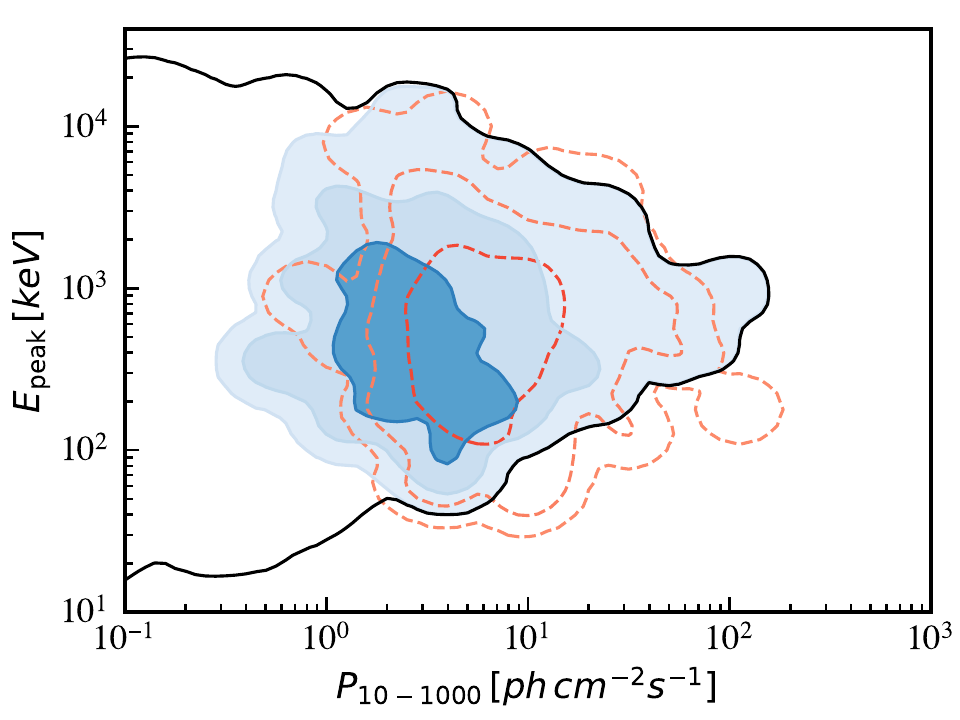}
}
\caption{Distribution of the GRB detected by HERMES (for the case of Constellation B, blue contours) in the plane $E_{\rm peak}$-Peak Flux (10-1000\,keV) compared with the distribution of real Fermi/GBM GRBs (red dashed contours). Contour levels show the regions containing 68\%, 95\%, and 99\% of the GRBs. Green contours show the HERMES GRBs detected only by the X mode. The black contour refers to the whole simulated population. Left: long GRB. Right: short GRBs.}
\label{fig:ep-flux}
\end{figure*}

We estimate that HP should detect $\sim$\,$19^{+5}_{-3}$\,yr$^{-1}$ short GRBs. All these are detected by the S instrument owing to the harder spectrum of short GRBs \citep{ghirlanda04,ghirlanda09}. Compared to {\it Fermi}-GBM ($\sim$ 40 short GRBs yr$^{-1}$, \citealt{poolakkil21}), the smaller detection rate of HP is due to its smaller effective area. HP should detect $\sim$\,$195^{+22}_{-21}$ yr$^{-1}$ long GRBs. About 26\% are detected thanks to the extension of the sensitivity to the soft X-ray energy range (3--20\,keV) through the X-mode (see Fig.~\ref{fig:ep-flux}). The latter, owing to the possible implementation of trigger algorithms based on the burst fluence, makes HP competitive with {\it Fermi}/GBM, which detects on average 200 yr$^{-1}$ long GRBs (\citealt{poolakkil21}).

We estimate the possible long and short GRB trigger rates by SpIRIT by assuming a scaling factor of 3/4 for the effective area with respect to that assumed for HP and a background count rate scaled for the same factor. This scaling factor is due to the instrument being delivered with one out of four non-functional quadrants. We also consider a 50\% observation efficiency given the SSO orbit. We find that SpIRIT could detect 84 (8) yr$^{-1}$ Long (Short) GRBs.

Concerning Constellation A, we estimate $\sim$\,$1565^{+292}_{-244}$ long and $\sim$\,$191^{+40}_{-28}$ short GRBs detected per year. 
For Constellation B, we find about $2468^{+531}_{-452}$ long GRBs and $327^{+78}_{-61}$ short GRBs per year.

Finally, our results show that constellations A and B could detect $\sim$\,38 and $\sim$\,70 long GRBs per year at $z>6$, mainly thanks to the X-mode low energy extension. Already with HERMES Pathfinder, we estimate that $\sim$\,3.4 GRBs at $z>6$ should be detected every year.  

\begin{figure*}
\centering
\includegraphics[width=18cm]{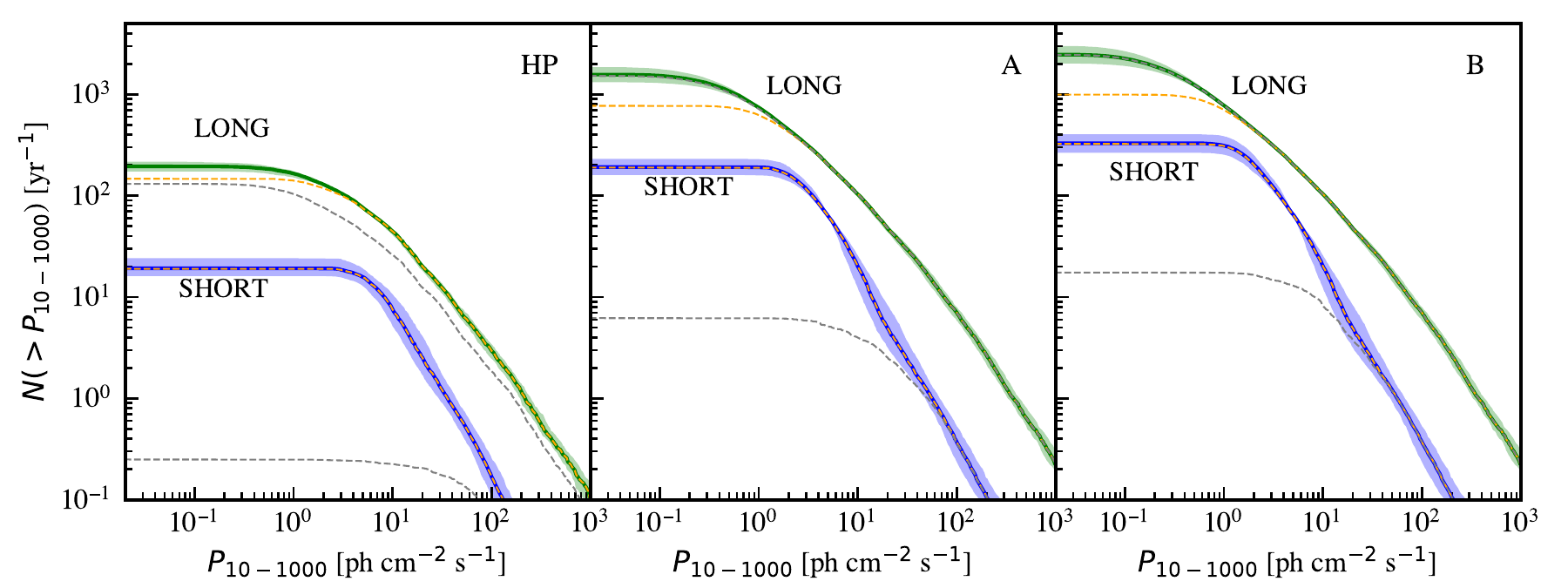}
\caption{Cumulative distributions of the photon peak flux (in the energy range 10-1000\,keV) for the GRBs detected by HERMES HP (left panel), Constellation A (center), and Constellation B (right). In each panel, the cumulative curve for long (short) GRBs is shown in green (blue), and the shaded region includes 68\% of the realisations.  Orange (gray) dashed curves show the contribution of the S (X) detection mode. For Constellation A and B, almost all long GRBs are detected by the X mode, making the dashed gray curve almost overlap with the total distribution (solid green curve).}
\label{fig:logN_logP}
\end{figure*}

\begin{figure*}
\centering
\includegraphics[width=18.cm]{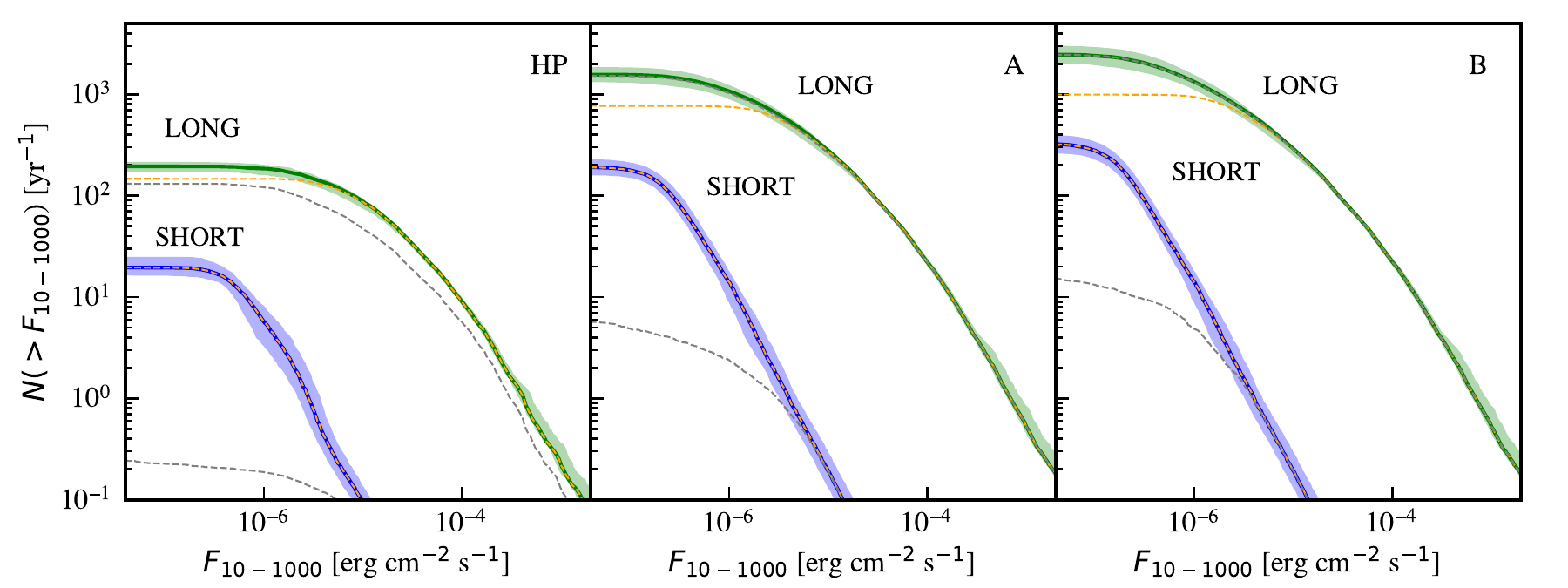}
\caption{Cumulative distributions of the fluence (in the energy range 10-1000\,keV) for the GRBs detected by HERMES HP (left panel), Constellation A (center), and Constellation B (right). In each panel, the cumulative curve for long (short) GRBs is shown in green (blue), and the shaded region includes 68\% of the realisations.  Orange (gray) dashed curves show the contribution of the S (X) detection mode.}
\label{fig:logN_logF}
\end{figure*}

The cumulative detection rates and their 68\% credible interval as a function of the photon peak flux and fluence (both integrated into the 10--1000\,keV energy range) are shown in Fig.~\ref{fig:logN_logP} and Fig.~\ref{fig:logN_logF}, respectively. As expected, a GRB with large fluence or falling within the FoV of both X and S (i.e., if it is at an angle $\theta<60^\circ$) is detected by both modes. The larger detection rate of the S-mode apparent at large flux/fluence is entirely due to the larger FoV of the S instrument.  
Moving to lower flux/fluence, the fraction of GRBs detected by the X-mode over the total increases. This is due to fainter events being also typically softer (i.e., their peak energy is located at lower energies), owing to the assumption of the $E_{\rm peak}$--$E_{\rm iso}$($L_{\rm iso}$) correlations\footnote{These correlations were implemented with their scatter in the population model, see \citealt{ghirlanda15}.} holding for long GRBs, thus favouring detections with the X-mode (X-only events in Table~\ref{tab:detection_rates}). 
This effect is clearly visible in Fig.~\ref{fig:ep-flux}, where the location of the HERMES GRBs in the observer frame $E_{\rm peak}$--peak flux plane is shown and compared to Fermi/GBM GRBs. Blue regions show the GRBs detected by HERMES-Constellation B and green contours mark the subsample of GRBs detected only by the X-mode. Red dashed contours show the GBM GRBs for comparison. Long HERMES GRBs (left-hand panel in Fig.~\ref{fig:ep-flux}) extend to lower peak energies, mostly thanks to the X detection mode, which allows the detection of softer GRBs compared to the GBM.  Our long GRB population model includes X--ray flashes (XRF) and X--ray rich GRBs (XRR) as a continuous extension of the luminosity function to low values (e.g., based on the result of \citealt{pescalli15}) together with the assumption of the  $E_{\rm peak}-E_{\rm iso}$ correlation. Therefore, long GRBs, XRF, and XRR are considered to be a unique population. If XRF and/or XRR  had an intrinsic rate exceeding that of "hard" GRBs, the detection rates estimated in this work would increase. Therefore, the soft-energy extension of HERMES detectors down to 4 keV represents an opportunity to explore the nature and rate of XRF and XRR bursts (see Fig.~\ref{fig:ep-flux}). Since short GRBs are, on average, harder than long ones, the short GRB population accessible to HERMES and to the GBM are more similar (right-hand panel in Fig.~\ref{fig:ep-flux}).

Note that the $E_{\rm peak}-E_{\rm iso}$ correlation is assumed to hold among long GRBs for the construction \citep{ghirlanda15} of the synthetic population used here. However, there are two outliers of this correlation, namely GRB980425 and GRB031203 (but see \citealt{Ghisellini2006}), which might be representative of a larger number of similar events \citep{Heussaff2013}. These appear as relatively low luminosity bursts ($\sim 10^{47-48}$ erg s$^{-1}$) with large intrinsic peak energies (in the range $100-250$ keV). The observed fluence of GRB980425 is $\sim 3\times 10^{-6}$ erg cm$^{-2}$ which would make it detectable by HERMES Constellation A and B (mid and right panels of Fig.\ref{fig:logN_logF}) while at the limit of the performances of HERMES-Pathfinder. The intrinsic rate density of  980425-like low luminosity events is$\rho_{LL}\sim 230^{+490}_{-190}$ Gpc$^{-3}$ yr$^{-1}$ \citep{Soderberg2006,pescalli15}. By computing the maximum distance out to which a GRB980145-like event would be detectable with a significance of $>5 \sigma$ by HERMES, we estimate that HERMES Pathfinder would detect a GRB980425--like event every $\sim$ 5 years. HERMES Constellation A (B) would be able to catch $\sim$5 (10) yr$^{-1}$ of such events owing to its higher sensitivity and full sky coverage.

To quantify a representative detection threshold of HERMES Pathfinder for long and short GRBs, we estimate for which limiting values of the peak flux the fraction of detected events is $\ge$80\% of the total simulated events (falling within the FoV). This value is reported in Table~\ref{tab:threshold}.

\begin{table*}[ht]
    \centering
    \begin{tabular}{|c||c|c|c|c||c|c|c|c|}
    \hline
               & \multicolumn{4}{c||}{LONG}& \multicolumn{4}{c|}{SHORT} \\
 \hline
               & Total &S& X& X-only    &Total &S& X& X-only    \\
    \hline
     HP & $195^{+22}_{-21}$  & $147^{+13}_{-11}$ & $131^{+17}_{-19}$ & $51^{+12}_{-13}$ 
     & $19^{+5}_{-3}$ & $19^{+5}_{-3}$& $0.3^{+0.4}_{-0.2}$ & $<0.1$    \\
    \hline
    Const. A  & $1565^{+292}_{-244}$ & $772^{+67}_{-61}$ & $1517^{+289}_{-251}$ & $796^{+221}_{-204}$ & $191^{+40}_{-28}$ & $188^{+38}_{-28}$ & $7^{+5}_{-4}$ &  $<3$  \\
    \hline
    Const. B  & $2468^{+531}_{-452}$ & $996^{+119}_{-64}$ & $2455^{+529}_{-460}$ & $1473^{+419}_{-375}$ &  $327^{+78}_{-61}$ & $322^{+76}_{-58}$ & $18^{+8}_{-10}$ & $<7$  \\
    \hline
    \end{tabular}
    \vskip 0.2truecm
    \caption{GRB detection rates (number of events per year) expected with HERMES (HP and Constellation) for the short and long classes. For each GRB class, the table reports the total rate and the partial rates on the single instruments (S - all events detected by S; X - all events detected by X; X-only - all events detected only by the X instrument). The difference in configurations A and B for the Constellation is the FoV of the single satellite (see Table~\ref{tab:instrument_parameters}).}
    \label{tab:detection_rates}
\end{table*}

\begin{table*}[ht]
    \centering
    \begin{tabular}{|c||c|c||c|c|}
    \hline
        & \multicolumn{2}{c||}{LONG}& \multicolumn{2}{c|}{SHORT} \\
 \hline
        & $P_{\rm lim}^{50-300}$ &$P_{\rm lim}^{10-1000}$ &  $P_{\rm lim}^{50-300}$ &$P_{\rm lim}^{10-1000}$   \\&ph\,cm$^{-2}$\,s$^{-1}$&ph\,cm$^{-2}$\,s$^{-1}$&ph\,cm$^{-2}$\,s$^{-1}$&ph\,cm$^{-2}$\,s$^{-1}$\\
        \hline
    HP & 0.8  & 3.1 &   5.6 & 12.1    \\
     \hline
    Const. A & 0.07 & 0.41   & 1.1  & 2.4   \\
    \hline
    Const. B & 0.02 & 0.15  & 0.74  & 1.6   \\ 
    \hline
    \end{tabular}
    \vskip 0.2truecm
    \caption{Sensitivity of HERMES (Pathfinder and constellation configurations A and B) in terms of detection threshold photon flux (at 1\,s for long and 64\,ms for short GRBs) in two different energy ranges: 50-300\,keV and 10-1000\,keV. The sensitivity is calculated as the value above which at least 80\% of the GRBs falling within the FoV are actually detected at more than 5$\sigma$. For each single module of the HP and of the Constellation A and B, the FoV is reported in Table~\ref{tab:instrument_parameters}. 
    }
    \label{tab:threshold}
\end{table*}


\section{Detection of short GRBs as gravitational waves counterparts}

The first, and still unique, event with an associated electromagnetic (EM) counterpart, GW170817 \citep{Abbott2017ApJ}, marked the beginning of the MM era. Produced by the merger of two neutron stars, this event was associated with the short GRB~170817A \citep{Abbott2017ApJ} and, thanks to intensive worldwide multi-wavelength follow-up efforts, to the optical near-infrared transient AT2017gfo \citep{coulter17}, spectroscopically classified \citep[e.g.][]{pian17} as a kilonova (KN, \citealt{Li1998,Metzger2019}). The study of the thermal (KN) and non-thermal (short GRB and afterglow) emission of GW170817 provided us with unique clues on the dynamics and ejecta properties resulting from a BNS merger, the formation sites of heavy elements, the progenitors of short GRBs (see e.g.\ \citealt{Nakar2021,margutti21}).

GRBs are narrowly collimated sources (e.g. \citealt{frail01}) with highly relativistic outflows such that their brightness, and consequently their detection probability, depends, in addition to their luminosity and distance, also on the orientation of their jets in the plane of the sky. Therefore EM detectors, given their limiting sensitivity, introduce an instrumental selection bias on the jet orientation, with more aligned jets being detected at larger distances. Similarly, the GW signal intensity depends, in addition to the source distance, also on the orientation of the binary orbital plane with respect to the line of sight: face-on events (with their orbital plane perpendicular to the line of sight) are detectable by GW interferometers at larger distances (setting the GW network horizon). By symmetry arguments, the GRB jet axis is expected to be perpendicular to the binary orbital plane, and hence the GW and GRB brightness are correlated.

The current GW detector network can detect BNS mergers out to a few hundred Mpc. Based on present BNS detections, the inferred local rate density of BNS mergers is between $10$ and $1700$ events per Gpc$^{3}$ per yr \citep[][union of 90\% credible intervals from different methods]{Abbott2023_GWTC3pop}.  In the fourth LVK observing run (O4, started in May 2023), despite the expected ~10 yr$^{-1}$ GW detections of BNS, only up to one joint GW-GRB detection is envisaged \citep{colombo22}. Such a small rate is mainly due to the limited GW horizon, within which, given the small rate density of SGRBs, detected events will be dominated by jets seen off-axis. In turn, these would be too faint for the current GRB detector performances. 

Third-generation gravitational wave interferometers such as the ET, expected to be operational in 2035 \citep{maggiore20}, will extend the GW horizon out to several Gpc, thus including the full BNS cosmic population up to and beyond the redshift of the peak of the BNS merger rate density. ET will detect $\mathcal{O}(5)$  BNS mergers per year, of which a considerable number $\mathcal{O}(1-2)$ will be sufficiently bright to be jointly detected in $\gamma$-rays by future space-based missions as prompt short GRBs 
\citep{Ronchini22}.

\subsection{Joint GW-EM detection rate}

\begin{figure*}
    \centering
    \includegraphics[width=\textwidth]{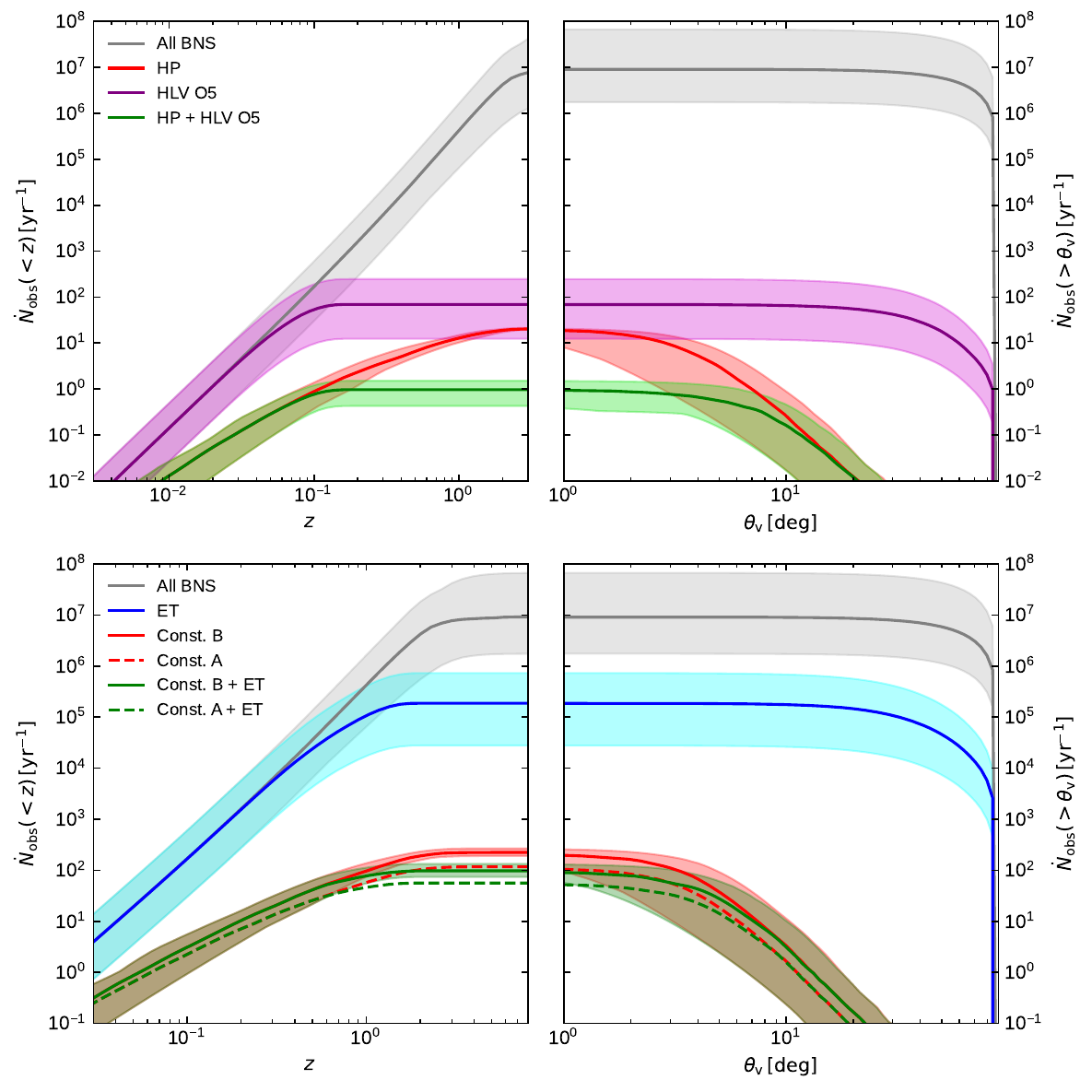}
    \caption{\textit{Top-left panel:} cumulative detection rates as a function of redshift at the time of the O5 GW detector network run. Different colours refer to sub-populations that satisfy different detection cut combinations (grey: all events; purple: detection by the HLV GW network with O5 sensitivity; red: detection by HERMES Pathfinder; green: joint detection by HERMES and the HLV network). For each sub-population, the solid line shows the median cumulative rate, while the shaded band encompasses the 90\% credible range at fixed $z$. \textit{Top-right panel:} inverse-cumulative rates of events with a viewing angle larger than a given threshold. The meaning of colours, lines, and bands is the same as in the top-left panel. \textit{Bottom-left panel:} similar to the top-left panel, but for HERMES constellations A and B, together with the ET GW detector at design sensitivity (ET-D). Light blue refers to the ET cumulative detection rate; the red solid line and shaded region show the detection rate (median and 90\% uncertainty) by the HERMES Constellation B, while the red dashed line shows the median for Constellation A. The green solid line and green shaded region refer to events detected by both ET and HERMES Constellation B. The dashed line shows the median detection rate for ET and HERMES Constellation A. \textit{Bottom-right panel:} similar to the top-right panel, but for ET and HERMES Constellations A and B.}
    \label{fig:GWEM_new}
\end{figure*}

We estimated the joint GW-EM detection rates adopting the same structured jet short GRB population as in the previous sections \citep{Salafia2023}. We assumed all short GRBs to be produced in the aftermath of a binary neutron star (BNS) merger. Hence, we considered only the population model hyperparameter values\footnote{We used the hyper-posterior samples for the fiducial `full-sample analysis' in \citet{Salafia2023}, publicly available at \url{https://doi.org/10.5281/zenodo.8160783}} that yield a local rate density of SGRBs (including all viewing angles) consistent with the uncertainty range $10$ -- $1700\,\mathrm{Gpc^{-3}\,yr^{-1}}$ estimated for BNS mergers from the GWTC-3 catalog population analysis \citep{Abbott2023_GWTC3pop}.

Given the expected evolution of the HERMES project, we considered HP for the prediction of the joint GW plus GRB detection rates expected during the fifth observing run (O5) of the current ground-based GW detector network (we considered LIGO Handford, LIGO Livingston, and Virgo, namely a `HLV' network). For predictions relevant to the era of next-generation ground-based GW detectors, we considered HERMES Constellations A and B and took the Einstein Telescope (ET) single triangle 10--km arm configuration as the representative GW detector. 

In order to estimate the GW detection rates, we considered a chirp mass 
$M_\mathrm{chirp}=1.18\,\mathrm{M_\odot}$ (similar to GW170817) for all BNS, we adopted the ET-D sensitivity curve \citep{ETsensitivity} for ET, and the projected O5 sensitivity curves \citep{O5sensitivity} for the HLV network. We assumed 100\% duty cycles for simplicity. We computed the signal-to-noise ratio (S/N) in each GW detector using the simple \texttt{gwsnr} code\footnote{\url{https://github.com/omsharansalafia/gwsnr}}, employing the quasi-Newtonian inspiral frequency-domain waveform from \citet{Maggiore2018}, with a cut-off at the innermost stable circular orbit frequency. We considered a detection whenever the network signal-to-noise ratio S/N$_\mathrm{net}$ (i.e.\ the sum square of the S/N of each detector in the network) exceeded S/N$_\mathrm{net}>12$. The capability of HERMES to detect the GRB counterpart was modelled as described in the previous sections. 

The resulting cumulative detection rates are shown in Fig.\ \ref{fig:GWEM_new}. The top panels refer to O5, while the bottom panels refer to the ET era. The left-hand panels show cumulative detection rates within redshift $z$, while the right-hand panels show cumulative detection rates for viewing angles larger than $\theta_\mathrm{v}$. The latter is defined as the angle between the GRB jet axis (assumed perpendicular to the binary orbital plane) and the line of sight.  In each panel, the full population (as observed with a hypothetical infinitely sensitive detector) is shown in light grey, while other colors refer to detection by HERMES (red), the GW detector network (purple for the HLV network, cyan for ET), or both (green). The shaded band around each line encompasses the 90\% credible range of the corresponding rate, as derived from the posterior probability on the hyper-parameters of the \citet{Salafia2023} model. 

Table \ref{tab:GWEM} reports the joint detection rates, including also the less likely configurations of HERMES HP together with ET, which would imply an unsuccessful upgrade of the detector sensitivity over the next decade, and the combination of Constellations A and B with O5, which would instead imply a very fast development of the Constellations.

\begin{table}
    \caption{Joint GW-EM detection rates. Number of events per year of GW BNS events detected during HLV O5 and by Einstein Telescope (ET) with their prompt jet emission (GRB) detected by HERMES.}
  
\centering
    \begin{tabular}{c|c|c|c}
         & HP & Const.A & Const.B \\
 \hline
 O5     &   $0.97_{-0.54}^{+0.54}$    &   $2.5_{-1.7}^{+1.8}$        &    $3.5_{-2.4}^{+2.7}$         \\
 ET     &   $12_{-2}^{+2}$   &    $56_{-12}^{+7}$          &   $97_{-23}^{+36}$         \\
    \end{tabular}
  \label{tab:GWEM}
\end{table}

\section{Discussion and Conclusions}

We have computed the rates of long and short GRBs detectable by HERMES. We have considered the HERMES Pathfinder instrument set up and two possible improved configurations, namely the HERMES Constellations (see Table~\ref{tab:instrument_parameters} and Fig.~\ref{fig:eff_areas}). In doing this, we have accounted for the effective area boresight angle dependence and for the sky coverage of the possible HERMES configurations, adopting state-of-the-art population models for short and long GRBs.

We find that (Table~\ref{tab:detection_rates}) $\sim$195 ($\sim$19) yr$^{-1}$ long (short) GRBs should be detected by the HERMES Pathfinder (about 80 long GRBs and 8 short GRBs per year should be detected by SpIRIT). This assumes an observation efficiency of 50\%, which is appropriate for the SSO of SpIRIT, and it is smaller than the 75-80\% achievable on equatorial LEO orbits.

A substantial increase of the detection rates of long and short GRBs (Table~\ref{tab:detection_rates}) is expected for HERMES Constellation (configurations A or B) which assumes a $\times$4 larger effective area and nine CubeSats (Constellation A) and a possible reduction of the background (through a reduced field of view for the single CubeSat but a larger number of 13 units - Constellation B). Intriguingly, we estimate that $\sim$40-70 long GRBs per year at redshift $z>6$ can be detected by HERMES Constellation, thanks to its energy extension down to the soft X--rays. Considering the present sample of 8 GRBs known at $z>6$, collected over the last two decades, HERMES Pathfinder, with 3.4 yr$^{-1}$ at $z>6$, could already help double this sample in a couple of years, if NIR follow up of these events is secured by ground-based facilities. One of the major limitations to this goal is the precise localisation of the burst. Localisations obtained with the triangulation technique are limited by three main factors: 1) the projected baseline, 2) the temporal structure of the GRB, and 3) the GRB brightness (see e.g., \citealt{sanna2020}, \citealt{Sanna2023}, \citealt{Burgess2018}). HP and SpIRIT use a Low Earth orbit, implying a maximum projected baseline $\lesssim 10000-12000$~km. 
With a projected baseline of the order of 7000~km, we predict to be able, with the HP, to constrain at least one coordinate of bright bursts with temporal structure down to a few ms, with an accuracy ranging from a few tenths to several tens of degrees. The cause of poor sensitivity in one coordinate originates from the geometrical limitation due to the fact that the HP six units will fly on the same orbital plane. Constellations A and B would use at least two different orbital planes, allowing similar constraints on both GRB coordinates. Localisations accuracy would be significantly better than that achievable by HP due to the larger collecting area and the higher number of satellites. localisation accuracy of faint bursts with smooth temporal structure will be, in any case, limited using the triangulation technique with satellites in low Earth orbit (LEO). However, the localisation capabilities can be improved by joining the triangulation technique to other techniques, such as the counts aspect ratio technique used, for example, by the Fermi/GBM instrument, and by increasing the projected baseline. To the latter purpose, future constellations based on the HP concept (low cost, fast track) would greatly benefit from the infrastructure that is being set up around the Moon in the context of the Artemis program. Relay communications, navigation services, and accurate real-time positioning are all goals of the ESA Moonlight initiative\footnote{\url{https://www.esa.int/Applications/Connectivity\_and\_Secure\_Communications/Moonlight}}, which would enable the deployment of segments of GRB dedicated constellations around the Moon, thus improving localisation capabilities by a factor $\sim$50.

We have estimated the expected contribution of HERMES to detect the prompt emission of short GRBs as EM counterparts of GW events due to BNS mergers. These results are shown in Fig.\ref{fig:GWEM_new} and reported in Table \ref{tab:GWEM}. First of all, both during O5 and in the ET era, the short GRB detection horizon is set by the GW detector rather than HERMES: the latter can efficiently detect events almost out to $z\sim 3$, while the HLV network is limited to $z\lesssim 0.2$ and ET to $z\lesssim 2$. According to this population model (which predicts a strong event density evolution with redshift, see \citealt{Salafia2023}), the total rate of BNS mergers in the Universe is very large, $\dot N_\mathrm{BNS,tot}=9.2_{-7.4}^{+58}\times 10^{6}\,\mathrm{yr^{-1}}$. The HLV network can detect around one in $10^5$ of these events, with a total predicted detection rate in O5 of $\dot N_\mathrm{obs,HLV}=70_{-57}^{178}\,\mathrm{yr^{-1}}$. Only a small fraction of these events is detected in conjunction with HERMES HP, $\dot N_\mathrm{obs,HP+HLV}=1.0\pm{0.5}\,\mathrm{yr^{-1}}$. The right-hand panels show that while the typical viewing angle of jets that produce an EM-detectable GRB is around a few degrees, the joint GW+GRB events have a larger typical viewing angle of several degrees.

The multi-messenger prospects improve significantly in the third-generation GW detector era: ET can detect more than 1\% of the total population, $\dot N_\mathrm{obs,ET}=1.9_{-1.6}^{+5.4}\times 10^{5}\,\mathrm{yr^{-1}}$. Thanks also to the improved HERMES Constellation sensitivity, the joint GW+GRB detection rates are also much more encouraging: for Constellation A, we predict $\dot N_\mathrm{obs,A+ET}=56_{-12}^{+7}\,\mathrm{yr^{-1}}$, while for Constellation B the joint detection rate increases to $\dot N_\mathrm{obs,B+ET}=97_{-23}^{+36}\,\mathrm{yr^{-1}}$. The typical viewing angle of BNS mergers observed as a multi-messenger source in the ET era, on the other hand, is closer to that of EM-only events: this is because the GRB brightness decreases quickly with the viewing angle \citep{Salafia2023} and hence the increased joint detection rate follows mainly from the improved GW horizon, while the improved GRB sensitivity of the HERMES Constellations does not push the limiting viewing angle to much larger values.

\begin{acknowledgements}
This research acknowledges support from the European Union Horizon 2020 Research and Innovation Framework Programme under grant agreements HERMES-SP n. 821896 and AHEAD2020 n. 871158, and by ASI INAF Accordo Attuativo n. 2018-10-HH.1.2020 HERMES—Technologic Pathfinder Attivit\`a scientifiche.  GG acknowledges financial support from the MUR, PRIN 2017, grant No. 20179ZF5KS, and the project GRB PrOmpt Emission Modular Simulator (POEMS) financed by INAF Grant 1.05.23.06.04. GG, LN and OS acknowledge funding by the European Union-Next Generation EU, PRIN 2022 RFF M4C21.1 (202298J7KT - PEACE). This research was partially supported by the Australian Research Council Centre of Excellence for All Sky Astrophysics in 3 Dimensions (ASTRO 3D), through project number CE170100013. The SpIRIT mission acknowledges support from the Australian Department of Industry Science and Resources grants ISI-EC000086 and MTMDM000034 for technical development and in-orbit operations.  A.T. acknowledges financial support from ASI-INAF Accordo Attuativo HERMES Pathfinder operazioni n. 2022-25-HH.0. 
\end{acknowledgements}

\appendix

\bibliographystyle{aa} 
\bibliography{biblio} 

\end{document}